\documentclass[12pt]{iopart}

\newcommand{\Journal}[4]{{#1} {\bf #2}, #3 (#4)}
\usepackage{graphicx} 
\usepackage{color}

\newcommand{\PLB}{{\em Phys. Lett.}  {\bf B}}

\newcommand{\PRC}{{\em Phys. Rev.} C}
\newcommand{\PRP}{{\em Phys. Rep. }}

\newcommand{\NIMA}{\em Nucl. Inst. Meth. A:}

\def\Journal#1#2#3#4{{#1} {\bf #2},{#3}{(#4)}}

\def\APJ{\em Astroph. J.}

\def\JCAP{\em JCAP}

\def\APP{\em Astroparticle Physics}

\def\JPG{\em Journal of Physics G}
\def\EPJC{\em Eur. Phys. J. C}

\def\NAT{\em Nature}

\def\ra{\rightarrow}
\def\be{\begin{equation}}
\def\ee{\end{equation}}

\newcommand{\nel}{\mbox{$\nu_e$}}

\newcommand{\bea}{\begin{equation} \begin{array}{c}}
\newcommand{\eea}{ \end{array} \end{equation}}

\newcommand{\xeho}{\mbox{$^{131}$Xe }}

\begin{document}

\title[Real-time measurements of solar $pp$ neutrinos using 
\xeho]{Real-time measurements of solar $pp$ neutrinos using \xeho}

\author{J Kostensalo and J. Suhonen}
\address{University of Jyv\"askyl\"a, Department of Physics, 40014 Jyv\"askyl\"a, Finland} 
\ead{joel.j.kostensalo@student.jyu.fi}

\author{K Zuber}
\address{Institut f\"ur Kern- und Teilchenphysik, TU Dresden,
Zellescher Weg 19, 01062 Dresden, Germany}
\ead{kai.zuber@tu-dresden.de}

\begin{abstract}
Various large-scale experiments for double beta decay or dark matter are based 
on xenon. Current experiments are on the tonne scale but there are also ideas to aim 
for even larger sizes in the future. Here we study the potential of the isotope 
\xeho to allow to make real-time measurements of
solar $pp$ neutrinos, besides classical neutrino-electron scattering. 
Improved nuclear models are used to determine the cross-section of
neutrinos on \xeho. The present calculations deviate significantly from the 
previous ones due to updated estimates for the excited-state contributions.
The updated capture-rate 
estimate for $pp$ neutrinos is $(4.47\pm 0.09)\ \rm SNU$ and for all 
solar neutrinos $(80\pm 21)\ \rm SNU$, with neutrino survival probabilities taken 
into account. Depending on the amount of Xe, solar $pp$ neutrinos might be measured
with rates of 100 per day, thus allowing to monitor them 
in real time for a long period.
\end{abstract}

\maketitle

\section{Introduction}
\label{sec:intro}
Neutrinos play a crucial role in modern particle, nuclear and astrophysics, 
including cosmology \cite{eji19,zub20}. 
It has been a major achievement of the last 25 years to show that neutrinos 
have a non-vanishing 
rest mass and that the problem of missing solar neutrinos could be 
solved \cite{ahm04}. 
Solar-neutrino measurements are one of the most important ingredients for 
the understanding of the Sun,
especially for the nuclear fusion processes. The low-energy region (below 1 MeV) 
of solar neutrinos has been observed in various 
radiochemical experiments starting from the famous Homestake
experiment \cite{cle98}, as well as further radiochemical experiments based 
on $^{71}$ Ga, namely 
GALLEX \cite{ham99}, GNO \cite{alt05,kae10} and SAGE \cite{abd09}, which were 
able to enter the energy region of the most abundant $pp$ neutrinos. All  
of them were radiochemical counting experiments 
without any energy or time information.
Hence, it is desirable to perform real-time experiments in the low-energy region.
Such measurements  have successfully been performed by Borexino over the last decade. 
Here the signal is the neutrino-electron scattering in a liquid scintillator.
Due to the extremely low impurity level,
Borexino was able to detect all $pp$-chain reactions \cite{bel18}, and very 
recently it also announced the first detection of
CNO neutrinos \cite{bel20}.   
Given the success of all these experiments, special dedicated solar-neutrino 
experiments will likely not be 
performed anymore. However, potential improvements on solar-neutrino data  
can still be made in a parasitic way. They are linked to  astro-particle 
and nuclear physics which - among others -
include topics like searches for dark matter or neutrinoless double beta 
decay. As neither of these two have been observed,
experiments have to push towards
larger and larger masses for a potential discovery. Thus, multi-ton-scale 
next-generation experiments, looking for double beta decay and dark matter, will 
appear. These experiments will be large enough to allow interference 
of solar neutrinos with the measurement signals. 
On the positive side, this might allow to gain  more information on solar neutrinos .\\

Here we want to explore the potential of 
the new large-scale Xenon-detector experiments \cite{bar14}. 
These kind of experiments will also detect
solar neutrinos via neutrino-electron scattering - like scintillators -  
producing recoil electrons. Here the focus for solar neutrinos
is on one particular Xenon isotope, \xeho, which has a threshold of 
355 $\pm 5$ keV \cite{aud17} for neutrino capture.
It is covering the highest 65 keV of the $pp$ neutrinos up to the 
endpoint of 420 keV. The natural  abundance of \xeho is 21.2\%.
Hence, an additional neutrino reaction, besides the electron scattering, 
is possible via
\be  
\nel + \,^{131}{\rm Xe}  \ra \,^{131}{\rm Cs} +e^-.
\ee
Unfortunately, no nice coincidence can be formed as the half-life of  the  
nuclide $^{131}$Cs is 9.69 days.
The detection signal of $^{131}$Cs would be X-rays of about 30 keV 
and Auger electrons around 3.4 and 24.6 keV. 
Such a long time spread prohibits a nice coincidence. The only way of detection
would be through the accumulation of $^{131}$Cs, combined with a sophisticated 
analysis of the low-energy spectrum. This might 
allow to identify a kind of peak by integrating the number of daughter decays. 


\section{Neutrino cross section}

\begin{figure}[ht]
\begin{center}
\includegraphics[width=0.65\columnwidth]{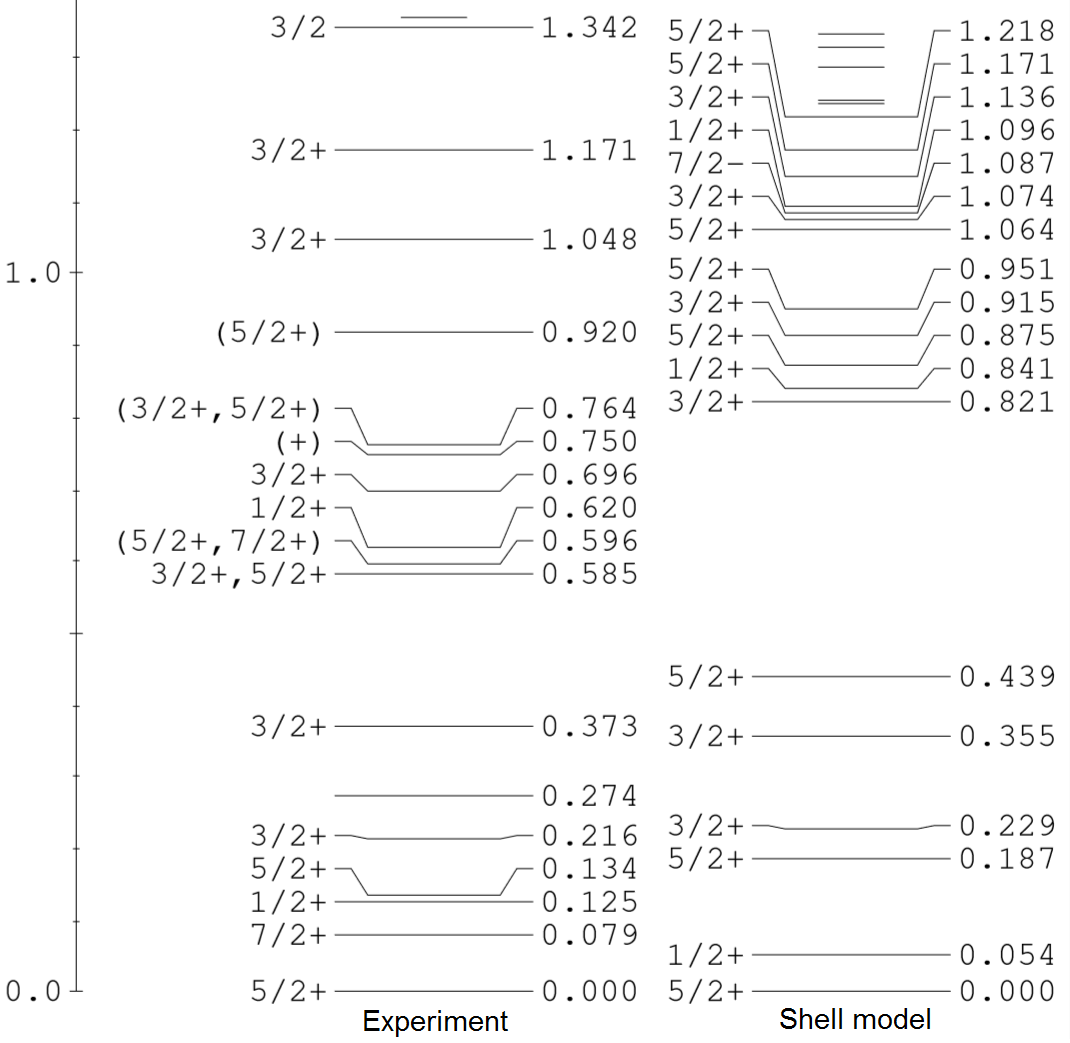}
\caption{Experimental and shell-model calculated energy spectra for $^{131}$Cs 
with only the states relevant for the neutrino-nucleus scattering off the 
$3/2^+$ ground state of $^{131}$Xe included.}
\label{fig:spec}
\end{center}
\end{figure}

\begin{figure}[ht]
\begin{center}
\includegraphics[width=0.65\columnwidth]{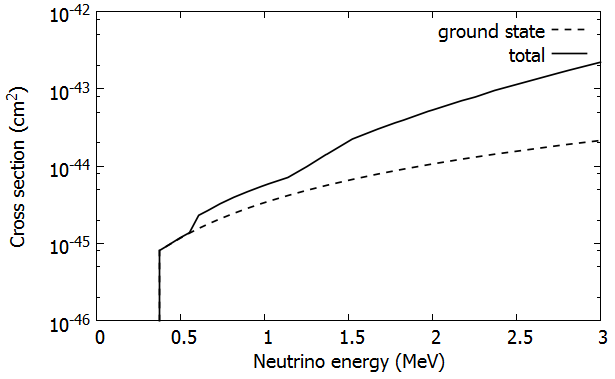}
\caption{Theoretical neutrino-nucleus cross section based on our shell-model 
calculation. The ground-state contribution can be related to the $^{131}$Cs 
half-life and is thus known to few tenths of per cent. The $pp$, $pep$, $^7$B, and 
CNO contributions are calculated using this cross section. 
The maximum energy of these neutrinos is about 1.7 MeV. }
\label{fig:shell}
\end{center}
\end{figure}

\begin{figure}[ht]
\begin{center}
\includegraphics[width=0.65\columnwidth]{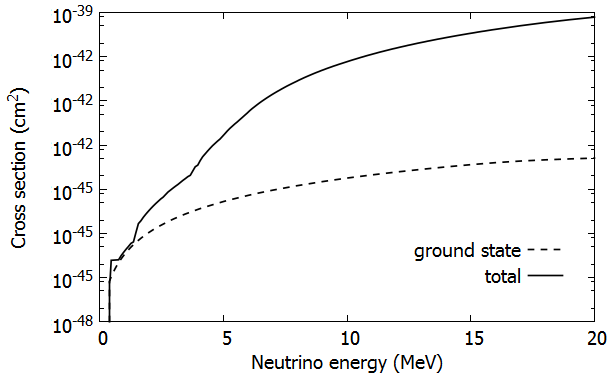}
\caption{Theoretical neutrino-nucleus cross section based on the microscopic 
quasiparticle-phonon model (MQPM). The $^8$B and $hep$ contributions are 
calculated using this cross section.}
\label{fig:mqpm}
\end{center}
\end{figure}

The neutrino-nucleus cross-section calculations presented here are based on
the Donnelly-Walecka method \cite{OConnell1972,Donnelly1979} for the treatment 
of semi-leptonic processes in nuclei. The necessary formulae and details 
of the application of the formalism can also be found in \cite{Ydrefors12}. 

Given that the threshold for the charged-current neutrino-nucleus scattering 
off $^{131}$Xe is 355$\pm 5$ keV, we need to calculate the final states up 
to about 1.4 MeV in order to cover the $pp$, $pep$, $^{7}$B, and CNO neutrinos. 
The initial and final states for these transitions, as well as the one-body transition 
densities, were calculated in the shell-model framework using the computer 
code NuShellX@MSU \cite{nushellx}. The calculations were done in a model 
space consisting of the orbitals $0g_{7/2}$, $1d$, $2s$ and $0h_{11/2}$, for 
both protons and neutrons, with the effective Hamiltonian 
sn100pn \cite{sn100pn}. Due to the huge computational burden of the 
shell-model calculations, some truncations had to be made. For the 
states $1/2^+_1$, $3/2^+_{1,2}$, and $5/2^+_{\rm g.s.,2}$ the neutron 
orbital $0f_{7/2}$ was filled with 8 neutrons, while no restrictions were 
posed on the other orbitals. However, the computational burden of this truncation was 
too large to allow calculations of more states. Instead, a second calculation was done 
keeping additionally the proton orbital $0h_{11/2}$ empty. This was done 
for 50 states of the spin 
parities $1/2^{\pm},3/2^{\pm},5/2^{\pm},7/2^{-}$ in $^{131}$Cs. The ground state 
of $^{131}$Xe was calculated in both cases with the same truncation 
as used for $^{131}$Cs in order to keep the calculations consistent. For the low energies 
considered here, the higher-multipole contributions are more than an order 
of magnitude smaller, and thus the higher-spin states are not relevant.

For $^{8}$B and $hep$ neutrinos we adopt the recent cross-section results 
of Pirinen et al. \cite{Pirinen2019}, which are calculated using the 
microscopic quasiparticle-phonon model (MQPM) \cite{mqpm1,mqpm2}. MQPM is an 
extension of the quasiparticle random-phase approximation to odd-mass 
nuclei. The MQPM approach can utilize a much larger model space than 
the shell model and thus for higher-energy neutrinos it is a good choice, 
while for the individual low-eneregy states the shell model is preferable.

The shell-model calculated energy spectrum of $^{131}$Cs is presented 
in figure \ref{fig:spec}. The shell model manages to predict the ground-state 
spin-parity correctly, as well as getting the density of states 
roughly right. Since the exact energies are important for the lowest energy 
states, we adopt the experimental energies for the excited states 
$1/2^+_1$, $3/2^+_{1,2}$ and $5/2^+_2$. Furthermore, the 
ground-state-to-ground-state scattering cross section can be deduced 
from the electron-capture half-life of $^{131}$Cs through the reduced transition
probability $B$(GT), obtained from
\begin{equation}
B(\textrm{GT})=
\frac{2J+1}{2J'+1}\frac{2\pi^3\hbar^7\mathrm{ln \  2}}{m_e^5c^4(G_{\rm F}
\cos \theta_{\rm C})^2}\times 10^{-\log ft} = 0.0178,
\end{equation}
where $J=5/2$ is the spin of the $^{131}$Cs ground state, $J'=3/2$ is the 
spin of the $^{131}$Xe ground state, $\theta_{\rm C}\approx 13.04^\circ$ is the 
Cabibbo angle, $G_{\rm F}$ is the Fermi coupling constant and $t$ is the
electron-capture half-life. The value of
the phase-space factor $f=5.548$ can be interpolated from the tables of 
Ref. \cite{gove}. For the rest of the 
states we adopt an effective value $g_{\rm A}=1.0$ of the axial-vector coupling 
in order to account for the limited model space 
(see e.g. the review \cite{Suhonen2017}).  

In the paper of Pirinen et al. \cite{Pirinen2019} the MQPM results for 
$^{8}$B neutrinos are given for $g_{\rm A}$ values of 0.7 and 1.0. 
We choose the value $g_{\rm A}=1.0$ here but fix the ground-state-to-ground-state 
cross section using the half-life of $^{131}$Cs, as was done in the case of 
the shell-model calculations.

The cross section as a function of the neutrino energy has been presented for 
the shell model in Fig. \ref{fig:shell} and for the MQPM in 
Fig. \ref{fig:mqpm}. For the energies relevant for $pp$, $pep$, $^{7}$B 
and CNO neutrinos, the cross section is dominated by the 
ground-state-to-ground-state contribution, while for $^{8}$B and $hep$ 
neutrinos the excited states dominate. 

The contributions for each component of the neutrino spectrum are given 
in Table~\ref{tbl:comps}. The largest contribution comes from $^{8}$B 
neutrinos. However, the $pp$ and $^{7}$B contributions are individually 
almost as large. The $^{8}$B contribution accounts for only about 36\% of 
the total expected events. Thus, calculating the solar-neutrino spectrum 
only for $^8$B neutrinos, as done in Ref. \cite{Pirinen2019}, results in a 
severe under prediction of the total number of events.  

\begin{table}
\centering
  \caption{The nuclear-structure model, survival probability and neutrino flux, 
adopted for each component of the solar-neutrino spectrum. The fluxes 
are from the solar model BS05(OP) \cite{Bahcall2005} and survival 
probabilities from \cite{Agostini2018}. The last two lines give the solar-neutrino
capture rates of the present work and Georgadze \textit{et al.} \cite{Klapdor}
in units of SNU. } 
\resizebox{\textwidth}{!}{
\begin{tabular}{@{}lcccccccc@{}}
\hline
& $pp$ & $pep$ & $hep$ & $^7$Be & $^8$B & $^{13}$N & $^{15}$O & $^{17}$F \\
\hline
Theory&SM&SM&MQPM&SM&MQPM&SM&SM&SM\\
Surv. & 0.54&0.5&0.36&0.54&0.36&0.54&0.5&0.5\\
Flux&5.99$\times10^{10}$&1.42$\times10^{8}$&7930&4.84$\times10^{9}$&
5.69$\times10^{6}$&3.07$\times10^{8}$&2.33$\times10^{8}$&5.84$\times10^{6}$\\
SNU (new)&4.47(9) &1.2(3)& 0.33(10)& 10.4(14) & 62(19)& 0.52(7)&0.9(2)& 0.024(5)\\
SNU \cite{Klapdor} & 5.2 & 0.80 & - & 9.6 & 4.6 & 0.86 & 0.90 & - \\
\hline
\hline
\end{tabular}}
\label{tbl:comps}
\end{table}

While the ground-state-to-ground-state contribution (which is 15.6 \% of the 
total cross section) is known very precisely, with an uncertainty estimated 
here as 2 \% due to the precision of the natural constants and the phase-space 
factor $f$, there 
are quite large uncertainties related to the other states. Assuming a 
conservative 30 \% uncertainty, independent of the ground-state error, for the 
rest of the states, the total solar-neutrino capture rate for $^{131}$Xe is
\begin{equation}
R (^{131}\rm Xe) = (80\pm 21)\  \rm SNU, \quad \quad (\mathrm{Solar \ neutrinos})
\end{equation}
which has been corrected for the survival probabilities from \cite{Agostini2018}, and
which is much larger than the old estimate 20.0 SNU \cite{Klapdor}. The 
$pp$ neutrino capture rate is predicted to be about 14 \% lower than the previous 
estimate. About 4 \% of this difference is explained by the updated 
interpolation of the $logft$-value, 0.5 \% by the updated flux in the new 
solar model, and the rest is due to the contribution of the excited states. 
For $^7$Be neutrinos the estimate for the rate is increased with respect to 
the previous results, owing to the increased $^7$Be neutrino-flux estimate in 
the more recent solar model \cite{Bahcall2005}. However, the results are in 
agreement within the uncertainties. The major difference in the rates 
relates to $^{8}$B neutrinos, which in our updated calculations are predicted 
to contribute over 10 times as many counts as in the old work. 
This difference highlights the 
vast uncertainties related to the excited-state contributions.

\section{Outlook and conclusion}
We have explored the potential for solar-neutrino measurements in the $pp$ region 
by using the isotope $^{131}$Xe.   
Large-scale Xe detectors are used since many years for searches of dark matter 
and neutrinoless double beta decay.
As nothing was found so far, experiments are upgraded and are now on the ton 
scale, like Xenon1T \cite{aal17} and LUX-ZEPLIN (LZ) \cite{ake20}. Even 
larger detectors are planned, like the DARWIN experiment using 50 tons 
of Xenon \cite{aal16}. 
Hence, such a $pp$ measurement based on \xeho might become reality. 

\section*{Acknowledgements}
This work has partially been supported by Academy of Finland under the
Academy project no. 318043. J. K. acknowledges the financial support from the Jenny and Antti Wihuri Foundation.
%
%

\section*{References}

\end{document}